\documentclass{llncs}
\usepackage{makeidx}  
\usepackage{amsmath}
\usepackage{graphicx}
\begin{document}
\frontmatter          
\pagestyle{headings}  
\addtocmark{Hamiltonian Mechanics} 
\mainmatter              
\title{Position-sensitive propagation of information on social media using social physics approach}
\titlerunning{Position-sensitive propagation}  
%
\author{Akira Ishii\inst{1} \and Takayuki Mizuno\inst{2}
Yasuko Kawahata\inst{3}}
\authorrunning{Akira Ishii et al.} 
%
\tocauthor{Akira Ishii, Takayuki Mizuno, Yasuko Kawahata}
\institute{Tottori University, Tottori 680-8552, Japan,\\
\email{ishii@damp.tottori-u.ac.jp},\\ 
\and
National Institute of Information,
Hitotsubashi, Chiyoda-ku\\
Tokyo 101-8430, Japan
\and
Graduate School of Information Science and Technology, The University of Tokyo, Hongo, Tokyo 113-8654 Japan}

\maketitle              

\begin{abstract}
The excitement and convergence of tweets on specific topics are well studied. However, by utilizing the position information of Tweet, it is also possible to analyze the position-sensitive tweet. In this research, we focus on bomb terrorist attacks and propose a method for separately analyzing the number of tweets at the place where the incident occurred, nearby, and far. We made measurements of position-sensitive tweets and suggested a theory to explain it. This theory is an extension of the mathematical model of the hit phenomenon.
\end{abstract}
\section{Introduction}
Research on the number of social media writes has been conducted by many researchers. Many researches on the transition of the number of tweets about topics that are raised and converged in a short period of time on specific topics such as movies or major incidents as well as analysis of contents. For example, statistical laws governing fluctuations in word use from word birth to word death was investigated by A. M. Petersen\cite{Petersen}. The propagation of informations were investigated for social events\cite{Liu}, financial markets\cite{Lillo}, and other human communications\cite{Gao,Tavares}.

Mathematical model of the hit phenomenon presented by Ishii et al. \cite{Ishii2012a} is one of the theories of social physics which analyzes the time course of such tweet number. This theory is used to analyze attention on Twitter and blogs about many topics.    In  several recent papers, it was shown that the theory is not only applicable to the  box office, but also other social entertainment such as local events\cite{Ishii2012b}, animated dramas on TV\cite{Ishii2013b}, the ``general election'' of the Japanese girl-group AKB48\cite{Ishii2013a}, online music\cite{Ishii2012c}, plays\cite{Kawahata2013d}, music concerts\cite{Kawahata2013a,Kawahata2013b}, Japanese stage actors\cite{Kawahata2013e}, Kabuki players of the 19th century\cite{Kawahata2013c},  TV dramas\cite{Ishii2014a}, and online games like Pokemon GO\cite{Ishii2016a}.

However, since the mathematical model of the hit phenomenon does not assume location information, it assumes analysis of the reputation in the world for any topic. However, there are cases in which there are differences in the way information is transmitted between those who witnessed on the spot like a bomb terror or football match and those who are not there. In that case, the tweet is different between the tweet of the witness and the person who reads the tweet and knows the information. The time difference between the tweets of those who witnessed and the tweets posted by the people who heard in the listening are considered interesting as research subjects.

In our previous study\cite{Ishii2012a}, it is assumed that the strength of interest of people attenuates exponentially. On the other hand, the presence of a power-law relaxation seems to be a common behavior in a wide range of complex systems\cite{Lillo-Mantegna2003}. Although it is known that this is known to occur in movies and the like , attention such as events and anniversaries is known to attenuate by a power function. \cite{Sano2013a,Sano2013b} In the case of social interest, we attenuate the intermediate between the exponential relaxation and the power-law relaxation \cite{Ishii-Koyabu}. 

For the position information of the tweet, we use the data attached to the tweet. On the other hand, the theory including the position information of tweet suggests extending the mathematical model of the hit phenomenon to the position sensitive sense, and presents a simple model calculation here.
\section{Twitter data at the incident}
Use the total of 2.8 million Tweets written on twitter with position information between 42.12303 and 42.58496 north latitude and 71.3099 west longitude to 70.8358 west latitude from 2012/12/20 to 2013/6/25. Tweet with location information is less than 1 percent of all tweets. 2013/4/15 14:45 (EST) Boston marathon bomb incident occurred at two points [42.349736, -71.078613] and [42.349148, -71.081179] at the same time. These points about 100 m apart are approximately at the center of the grit mentioned above. When the people $r$ km away from the place where the incident occurred got the information that the bomb exploded and how much it responded to that information, the unit of the word where the word after stemming is "bomb" Use frequency of occurrence per minute to investigate.

Focus on the tweet in the range from $r - dr$ km to $r + dr$ km from the center of the two places where the explosion occurred. Set the frequency of appearance in $r$ km at time $t$ of stemmed word "bomb" as $n (\vec{r}, t)$. Even before the incident, since its appearance frequency varies depending on the distance $r$, introduce the normalized appearance frequency number, as follows,
\begin{equation}
\overline{n}(r,t) = \frac{n(r,t)}{<n(r)>}.
\end{equation}

Here is the average number of occurrences of "bomb" per day at distance $r$, from 2012/12/20 to 2013/6/25 the day before the incident. Fig.1(a) is cumulative from 12/20/2012. Prior to the incident, it increased linearly to 1 in 1 day according to the Poisson process. However, it can be seen that the tendency to rise sharply from immediately after the incident.

\begin{figure}[!h]
\centering
\includegraphics[clip,width=12.0cm]{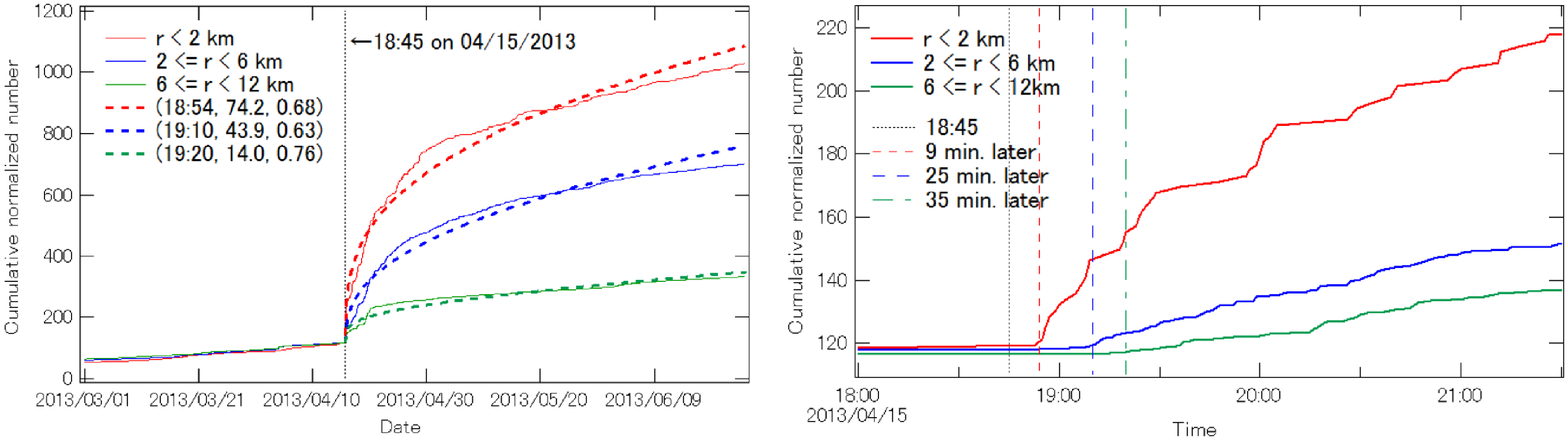}
\caption{Cumulative normalized number of “bomb” (a) from 03/01/2013 to 06/25/2013 and (b) from 18:00 to 21:30 on 04/15/2013. }
\label{fig:Exp}
\end{figure}

\begin{table}[htb]
\begin{center}
\caption{}
\begin{tabular}{lcrr}
Distance & Recognized time & Impact  $\alpha_r$ \hspace{1cm} & Reduction  $\beta_r$ \\
$r < 2km$	 & 18:54 &	$74.18 \pm 0.27$ \hspace{1cm}  & $0.683 \pm 0.003$ \\	
$2 \leq r < 6km$ & 19:10 & $43.87 \pm 0.13$ \hspace{1cm} & $0.627 \pm 0.003$ \\		
$6 \leq r  <12km$ & 19:20 & $13.95 \pm 0.11$ \hspace{1cm} & $0.760 \pm 0.004$
\end{tabular}
\end{center}
\end{table}

In order to clarify the dependence of the distance $r$ from the hypocenter at the time from the explosion to the tweet, we observe the time series of 3.5 hours before and after the case shown in Fig. 1 (b). It can be seen that the rapid rise after the incident is delayed depending on the distance $r$. This is because a gap occurs at the recognition time of the case between the person who directly witnessed the incident and the person who heard indirectly the incident occurred. Table 1 shows the recognition time at each distance. The greater the distance $r$, the better. The recognition time tends to be delayed.

Next, we will clarify to what extent people in distance $r$ respond to explosion information and lose interest according to what kind of process. Generally, it is known that the relaxation process of the frequency of appearance that has risen sharply after an event follows a power function with variables as the time of recognition or elapsed time since the start of the event. 
\begin{equation}
\overline{n}(r,t) =\alpha_r (t- \tau_r)^{-\beta_r} +c
\end{equation}
\begin{equation}
\overline{N}(r,t) = \frac{\alpha_r (t-\tau_r)^{1-\beta_r}}{1-\beta_r} +ct + constant
\label{mizuno}
\end{equation}

Here, $\alpha_r$ and $\beta_r$ represent the initial response to the event and the relaxation rate. This is because the occurrence frequency $c$ independent of the event is standardized. We investigate the dependence of the indices $\alpha_r$ and $\beta_r$ on the distance $r$. Table 1 shows the coefficient at each distance $r$, and Fig1. (a) shows the approximate line eq. (\ref{mizuno}) of those coefficients. It is found that the magnitude of the reaction remarkably decreases depending on the distance. In other words, interest in distant people's case is weak. On the other hand, representing the relaxation rate tends not to depend on the distance and shows a value in the vicinity of 0.7. This suggests that those who received a major shock on the incident and those who did not take it did not change the dynamics of shock relaxation much.

\section{Mathematical model for position-dependent social media information}

\subsection{mathematical model for the hit phenomenon}

 We write down the equation of purchase intention at the individual level  $I_i(t)$ as

\begin{equation}
\label{hitEq1}
\frac{dI_i(t)}{dt}=\sum_\zeta C_\zeta A_\zeta(t) + \sum_j D_{ij}I_j(t) + \sum_j \sum_k P_{ijk}I_j(t) I_k(t)
\end{equation}
where $D_{ij}$ and $P_{ijk}$ are the coefficient of the direct communication, the coefficient of the indirect communication\cite{Ishii2012a}.  The advertisement and publicity effects are include in $A_\zeta(t)$ which is treated as an external force. The index $\zeta$ means sum up of the multi media exposures.  
Word-of-mouth (WOM) represented by posts on social network systems like blog or twitter is used as observed data which can be compared with the calculated results of the model. The unit of time is a day. 

We consider the above equation for every consumers in the society. Taking the effect of direct communication, indirect communication, and the decline of audience into account, we obtain the above equation for the mathematical model for the hit phenomenon. Using the mean field approximation, we obtain the following equation as equation for averaged intention in the society. The derivation of the equation is explained in detail  in ref.\cite{Ishii2012a}. 

\begin{equation}
\frac{dI(t)}{dt}=\sum_\zeta C_\zeta A_\zeta(t) + D I(t) + P I^2(t) 
\end{equation}

If we include the exponential relaxation as the previous work\cite{Ishii2012a}, the equation become

\begin{equation}
\frac{dI(t)}{dt}=\sum_\zeta C_\zeta A_\zeta(t) - \alpha I(t)+ D I(t) + P I^2(t) 
\end{equation}

On the other hand, if we employ  the power-law relaxation, we obtain

\begin{equation}
\frac{dI(t)}{dt}=\sum_\zeta C_\zeta A_\zeta(t) - \frac{\beta}{(t-t_0)^{\beta+1}} I(t)+ D I(t) + P I^2(t) 
\end{equation}

Thus, as discussed in ref.\cite{Ishii-Koyabu}, including both the exponential and the power-law relaxation, we obtain the following equation.
\begin{equation}
\frac{dI(t)}{dt}=\sum_\zeta C_\zeta A_\zeta(t) - (\alpha + \frac{\beta}{(t-t_0)}) I(t)+ D I(t) + P I^2(t) 
\end{equation}

In this paper, we consider people's interest as a function of position rather than averaging in society as a whole. In other words, Tweets from places that are separated by the same distance from the incident are all averaged with the same tweet. Write down the original formula (\ref{hitEq1}) as a variable with only position and time as follows.

\begin{equation} 
\begin{split}
\frac{d I({\bf r},t)}{dt} = \int C({\bf r -r'},t)A({\bf r -r'},t) d{\bf r'} + \int D({\bf r -r'},t)I({\bf r'},t) d{\bf r'} \\
 + \int \int P({\bf r}, {\bf r' -r''},t)I({\bf r'},t) I({\bf r''},t) d{\bf r'}d{\bf r''}
\end{split}
\end{equation}

This equation averages people's conscious attention not by averaging in society as a whole but by a position. In other words, if it is the same distance from the place where the incident occurred, it is an approximation that the same content is thought. The relaxation effect discussed above is not included for simplicity, but it is easy to include it. 

Now, with this equation as the foundation, let us derive an equation to be solved by approximating the distance to two stages of short distance and long distance in order to analyze the tweet about a case like a bomb incident.
\begin{description}
\item{short distance} $I({\bf r},t) = I_{near}(t)$
\item{long distance} $I({\bf r},t) = I_{far}(t)$
\end{description}
The case is witnessed in the range of the short distance and the sighting information is successively tweeted. Let $A (t)$ be the sighting information directly watched. Meanwhile, the influence of other people 's tweets appears on Twitter' s timeline as well as short and long distances. Then, the short distance $I_ {near} (t)$ is as follows.
\begin{equation}
\frac{dI_{near}(t)}{dt}= C A(t) + D_{near} (I_{near}(t)+I_{far}(t)) + P_{near} (I_{near}(t)+I_{far}(t))^2
\end{equation}
On the other hand, if people in the long distance do not know what kind of incident is happening and even the sound of the incident can not be heard, the influence of the incident comes only from other people's tweets. Therefore, it becomes as follows.
\begin{equation}
\frac{dI_{far}(t)}{dt}= D_{far} (I_{near}(t)+I_{far}(t)) + P_{far} (I_{near}(t)+I_{far}(t))^2
\end{equation}
where we assume that the strength of the direct communication $D$ and the indirect communication $P$ are different depend on the distance as suggested by eq.(\ref{mizuno}).

Furthermore, considering a bomb incident etc., consider a model that divides the distance into three stages. In this case, the distance that the people can not perceive the turbulent atmosphere is defined as the middle distance, although people do not witness the incident directly, but the sound of sirens of police cars, fire trucks, or ambulances that hears the noise of an explosion . We distinguish medium distance from distant distance as follows.
\begin{description}
\item{Short distance} $I({\bf r},t) = I_{near}(t)$
\item{Medium distance} $I({\bf r},t) = I_{middle}(t)$
\item{Long distance} $I({\bf r},t) = I_{far}(t)$
\end{description}
The influence of the information directly witnessed is received only by the person of the short distance, and this is taken as $A_ {near} (t)$. Because the middle district people also know the turbulent atmosphere, suppose that the influence of that atmosphere is given by $A_ {middle} (t)$.。
\begin{equation}
\label{NearEq}
\begin{split}
\frac{dI_{near}(t)}{dt}= C_{near} A_{near}(t) + D_{near} (I_{near}(t)+I_{middle}(t)+I_{far}(t)) \\
+ P_{near} (I_{near}(t)+I_{middle}(t)+I_{far}(t))^2
\end{split}
\end{equation}
\begin{equation}
\label{MiddleEq}
\begin{split}
\frac{dI_{middle}(t)}{dt}= C_{middle} A_{middle}(t) + D_{middle} (I_{near}(t)+I_{middle}(t)+I_{far}(t)) \\
+ P_{middle} (I_{near}(t)+I_{middle}(t)+I_{far}(t))^2
\end{split}
\end{equation}
\begin{equation}
\label{FarEq}
\frac{dI_{far}(t)}{dt}= D_{far} (I_{near}(t)+I_{middle}(t)+I_{far}(t)) + P_{far} (I_{near}(t)+I_{middle}(t)+I_{far}(t))^2
\end{equation}
With these three simultaneous equations, it is possible to express how information is transmitted from a short distance to a long distance. There is nothing tweeted in advance in the information which is suddenly recognized by the explosion like a bomb terror attack. The bomb explosion and subsequent confusion will first increase the interest of short-distance people by $ A_ {near} (t) $. People at medium range also know about the incident to some extent at $ A_ {middle} (t) $, interest is evoked, interest rises somewhat later than those at short distance. Long distance people do not have direct information, so they know the incident for the first time at the stage when the tweets of short distance and middle distance people come out. Both are the stages until a message from the mass media is entered.

\section{Model Calculation}

We will try simple model calculation according to the equations (\ref{NearEq})(\ref{MiddleEq}) and (\ref{FarEq}). Since the incident is more interesting in the event scene, the strength $D_{near}$ of direct communication of the person close to the case is 0.3,  the middle distance $D_{middle}$ is 0.2, and the person far away is $D_{far}=0.1$.

The size of the sighting information was set to $ A_ {near} (t=0)=10 and  $ and $ A_ {middle} (t=0) = 1$. The size of the sighting information is large for people close to the distance and small for medium distance people. We set the sighting information after $t = 1$ to zero. 

\begin{figure}[!h]
\centering
\includegraphics[clip,width=9.0cm]{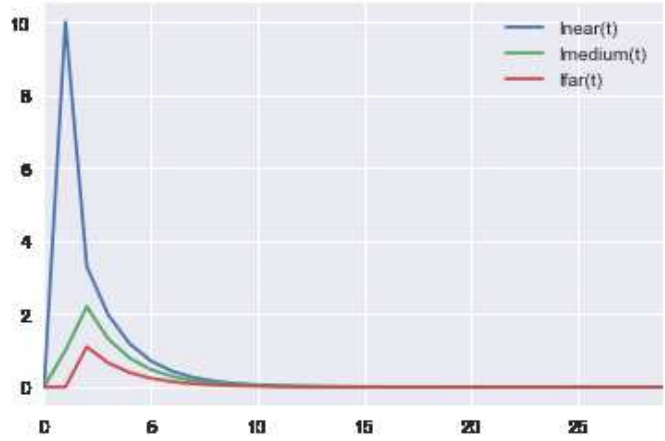}
\caption{Model calculation. $D_{near}=0.3$, $D_{middle}=0.2$, and $D_{far}=0.1$ are set and the indirect communication is neglected. $ A_ {near} (t=0)=10 and  $ and $ A_ {middle} (t=0) = 1$. }
\label{fig:calc1}
\end{figure}

Fig.\ref{fig:calc1} shows the calculation when the strength P of indirect communication is zero. From Fig. 1, it can be seen that it takes some time for the impact of the incident to be transmitted to persons at a middle distance or persons at a long distance. The impact of the incident fading is an exponential decay at any distance.

\begin{figure}[!h]
\centering
\includegraphics[clip,width=9.0cm]{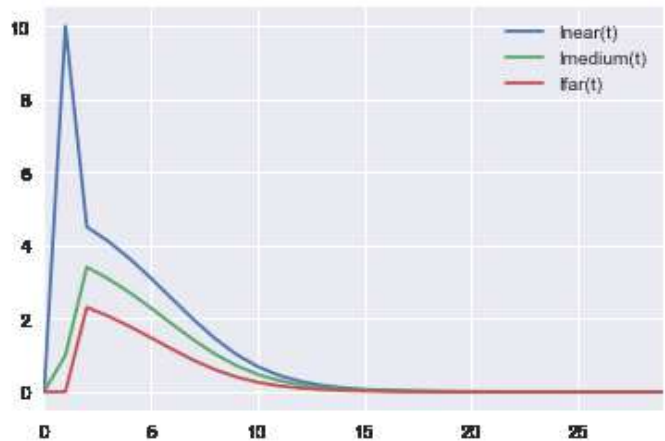}
\caption{Model calculation. $D_{near}=0.3$, $D_{middle}=0.2$, and $D_{far}=0.1$ are set and the indirect communication $P$ is set to be 0.01. $ A_ {near} (t=0)=10$  and $ A_ {middle} (t=0) = 1$. }
\label{fig:calc2}
\end{figure}

Fig. \ref{fig:calc2} shows calculation including indirect communication. $P = 0.01$ is set. From the calculation results shown in Fig. 2, it can be seen that the attenuation of information is considerably gentle by indirect communication. This effect seems to be remarkable especially for long distances. 

\begin{figure}[!h]
\centering
\includegraphics[clip,width=9.0cm]{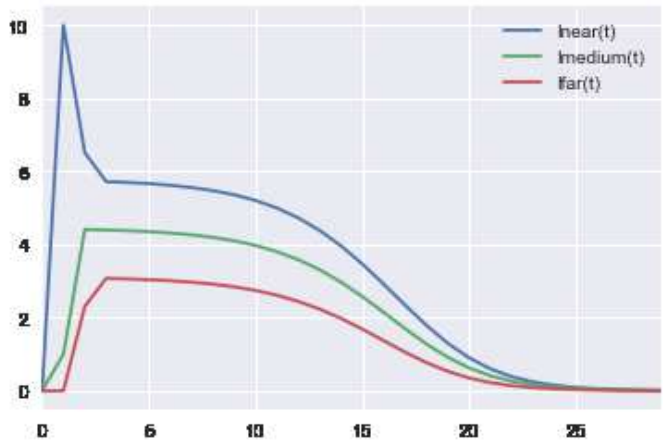}
\caption{Model calculation. $D_{near}=0.3$, $D_{middle}=0.2$, and $D_{far}=0.1$ are set and the indirect communication $P$ is set to be 0.01. $ A_ {near} (t=0)=10$,  $ A_ {near} (t=1)=1$, $ A_ {middle} (t=0) = 1$  and $ A_ {middle} (t=1) = 1$. }
\label{fig:calc3}
\end{figure}

Fig. \ref{fig:calc3} shows a calculation in which the sighting information of the incident continues to $t = 1$ after $t = 0$. We set $ A_ {near} (t=0)=10$,  $ A_ {near} (t=1)=1$, $ A_ {middle} (t=0) = 1$  and $ A_ {middle} (t=1) = 1$. It changes greatly in the calculation, and it is understood that the topic of the incident information continues for a longer time. 
In the case of shocking bombing terrorism, after the impact of the explosion, information comes into various people at short distance after the incident, such as the magnitude of the damage and the situation where police cars and ambulances rush. Therefore, in fact, this effect seems to make the reputation continue for a longer time.

Next, in order to compare with the measurement result in Fig.1, the calculation result is shown in the following fig.\ref{fig:calc4} and fig.\ref{fig:calc5} with cumulative distribution.Comparing the results with the observation results in Fig 1, fig.\ref{fig:calc5}  is better. Therefore, there seems to be indirect communication to some extent.

\begin{figure}[!h]
\centering
\includegraphics[clip,width=9.0cm]{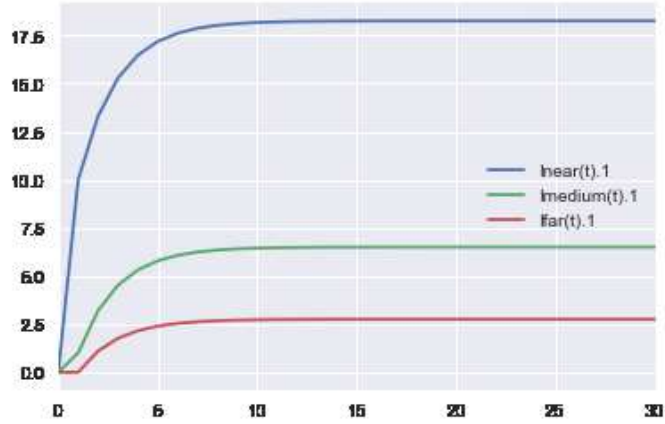}
\caption{ Cumulative distribution of calculation results. Ignoring indirect communication. .Sighting information $A (t)$ is only $t = 0$.}
\label{fig:calc4}
\end{figure}

\begin{figure}[!h]
\centering
\includegraphics[clip,width=9.0cm]{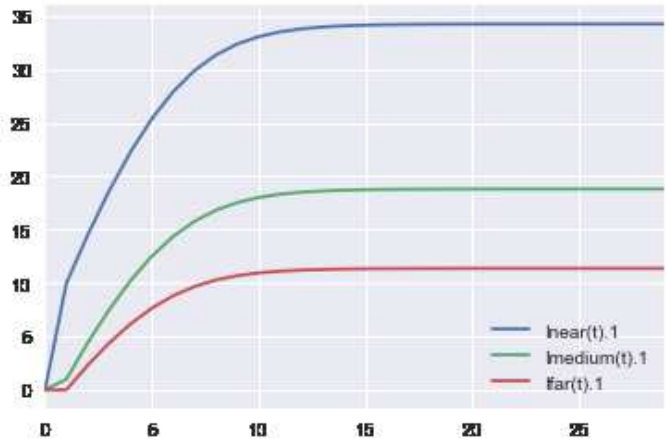}
\caption{ Cumulative distribution of calculation results. Including indirect communication. Sighting information $A (t)$ is only $t = 0$.}
\label{fig:calc5}
\end{figure}

\section{Discussion}

We analyzed tweets that talk about bombing terrorism cases using Tweet with location information. As a result, it was found that the number of tweets obviously depends on the distance from the incident site. This is clear from Fig.1. In the case of local incidents, the strength of interest also depends on the distance from the incident scene. From the measurement results of Tweet of this study, we can see that for local incidents it is necessary to analyze tweet by distance from incident scene. Also, by doing so, you can also see how the incident will be transmitted.

Therefore, in this research, we propose a theory extended with position dependency in mathematical model of hit phenomenon. As an approximate form of the theory, we derive a model that approximates the distance from the incident scene to three people, short distance people, middle distance people, and long distance people. In the theory proposed here, the strength $C$ of the reaction at the time of obtaining incident information, the strength $D$ of the influence by direct contact with others, and the strength $P$ of the indirect influence from the surrounding are important factors is there. From the analysis of tweet including position information, we considered that these coefficients have distance dependency. For the sake of simplicity in this research, distance dependence was adopted as follows.
As a model to be calculated, the reaction $CA (t)$ was set to $10$ for short range, $1$ for medium distance, $0$ for long range. The direct communication strength $D$ was set to 0.3 for short distance, 0.2 for medium distance, 0.1 for long distance.

Our model calculations on bombing terrorism cases are qualitatively in agreement with our measurements of tweet with positional information divided into three, short range, medium range, and long range. If we introduce distance dependence in more detail, we can analyze the propagation of local information from analysis of tweet with position information.

As described above, the mathematical model of the hit phenomenon that depends on the position information proposed by this research is a model that can analyze the number of tweets posted dependent on position information including distance dependency. By performing similar analysis on many examples, the distance dependence of the coefficients $C$ and $D$ becomes clear. It seems that it will become possible to quantitatively understand the manner in which the information of the large incident occurring locally propagates.

\section{Conclusion}

In this research, using the data of tweet with position information, we clarified the tweet of the terrorist bombing case, and made clear that the propagation of information has distance dependency. For the analysis, we introduced a mathematical model of hit phenomenon extended to form including position dependence. We obtained approximate equations divided into short distance, medium distance, and long distance from the incident scene. It was found that the model calculation by it can explain the measurement of the tweet with position information.


%
%

%

\end{document}